\documentclass[twocolumn,showpacs,preprintnumbers,amsmath,amssymb]{revtex4}

\usepackage{graphics,color,epsfig}
\usepackage{dcolumn}
\usepackage{bm}
\begin{document}
\topmargin .2cm

\title{ QUARK-GLUON-PLASMA FIREBALL EVOLUTION WITH ONE
LOOP CORRECTION IN THE PESHIER POTENTIAL }

\author{\bf S. Somorendro Singh\footnote{Email:sssingh@physics.du.ac.in }~ and R. Ramanathan }

%\email{sssingh@physics.du.ac.in}

\affiliation{Department of Physics and Astrophysics, University of Delhi, Delhi - 110007, India \\}

\begin{abstract}
  The study of free energy evolution of  
quark-gluon plasma (QGP)    
with one loop correction factor in the peshier potential is discussed.
The energy evolution with the effect of the correction factor
in peshier potential shows the transition temperature 
obtained in the range of temperature~$T=180-250~MeV$.
 The transition temperature is also effected with the decrease of
dynamical flow parameter of quark and gluon used in the potential
and it shows the 
observable QGP droplets 
of the stable size of fermi radius viz $2.5-4.5 fm$.   \\

\end{abstract}
                                                                                \vfill
\eject

\maketitle
 
\section{ \bf Introduction}

\large  The study of phase transition~[1]
from a confined matter of hadrons to a deconfined matter has become an
interesting topic in last two decades. During the early
stages of the universe formation,
 there was matter of deconfined
quarks and gluons, and in due process of cooling it leads to
matter of confined hadrons. The process of such early stage
of the universe is indeed explained
as a complicated phenomena by the heavy-ion
collider experiments. So, the study of quark-gluon plasma~(QGP)
fireball in Ultra Relativistic Heavy-Ion
Collisions becomes an exciting field in the present day of
heavy ion collider physics [2].  In this brief paper, we focus on the QGP
evolution through the free energy expansion of the system. To evolute the
free energy we use the peshier potential with one loop correction
to construct the density of states of particles in the system.
Thus the free energy evolution is obtained through this density of state.
Due to the correction factor in the potential through
coupling value~[3-5], there is a lot
of changes in the free energy expansion of QGP fireball and stability of
droplet with the variation of dynamical quark and gluon flow parameters.
\par In brief, the paper is organized as the construction of density of 
states with one loop correction in the potential and study the 
free energy evolution effected by the loop. In conclusion, we give the
details of evolution of QGP fireball with differently flow parametrization 
values of quark and gluon. 
 \section{Density of states for QGP 
 with one loop correction}
The potential $ V_{conf}(q) $ is
now modified with inclusion of one 
loop correction
factor from simple confining potential. The modified potential is therefore 
obtained through
the expansion of strong coupling constants of one loop factor within
the perturbation theory as~[6,7]:

\begin{equation}\label{3.18}
V_{\mbox{conf}}(q) = \frac{2 \pi}{q}\gamma ~ \alpha_{s}(q) T^{2} [1+\frac{\alpha(q)}{4\pi} a_{1} ]- \frac{m_{0}^{2}}{2 q},
\end{equation}

where
\begin{equation}\label{3.22}
\gamma = \sqrt{2}\times \sqrt{(1/\gamma_{g})^{2} + (1 / \gamma_{q})^{2}},
\end{equation}

called the effective rms value of parametrization factor 
with $ \gamma_{q}=1/8 $ and $\gamma_{g}=~(8-10)~\gamma_{q}$. 
These factors determine the dynamics of QGP flow and 
subsequent transformation to hadrons.~$ \alpha_{s}(q) $ is the coupling 
value of quark and gluon with degree of 
freedom~ $n_{f},$ as
\begin{equation}
 \alpha_{s}(q)=\frac{4 \pi}{(33-2n_{f})\ln(1+q^{2}/\Lambda^{2})},
\end{equation}
in which $~\Lambda $ QCD parameter is taken equal to~$ 0.15~$GeV.
The coefficient $a_{1}$  in the confining potential 
is the correction factor of one loop
connection in their interactions and it is given as~[8]:
\begin{equation}
a_{1}= 2.5833-0.2778~ n_{l},~ 
\end{equation}
where $n_{l}$ is considered with the number of light quark elements~[8,9].
 
 Now the density of states in phase space with loop correction  
in the peshier potential is obtained through Thomas and Fermi model as~[10]:

\begin{equation}
\int \rho_{q,g}dq=\nu/\pi^{2}[-V_{conf}(q)]^{2} \frac{dV_{conf}}{dq}~,
\end{equation}
 
or,

\begin{equation}\label{3.13}
\rho_{q, g} (q) =\frac{\nu}{\pi^2}[\frac{\gamma_{q,g}^{3}T^2}{2}]^{3} g^{6}(q)A, 
\end{equation}
where
\begin{eqnarray}
A=\lbrace 1+\frac{\alpha_{s}(q)a_{1}}{\pi}\rbrace^{2}[ \frac{(1+\alpha_{s}(q)a_{1}/\pi)}{q^{4}}+ \nonumber \\  
\frac{2 (1+2\alpha_{s}(q)a_{1}/\pi)}{q^{2}(q^2+\Lambda^2)\ln(1+\frac{q^2}{\Lambda^2})}] 
\end{eqnarray}
and~ $\nu$ is the volume occupied by the QGP and $q$ is 
the relativistic four-momentum in natural units 
and$~g^{2}(q)=4 \pi \alpha_{s}(q)$.    

\section{The free energy evolution}

The free energy 
of quarks and gluons is defined in the following with the density of states
as~[11]:
\begin{equation}\label{3.20}
F_i = \mp T g_i \int dq \rho_{q,g} (q) \ln (1 \pm e^{-(\sqrt{m_{i}^2 + q^2}) /T})~,
\end{equation}
 
with low energy cut off as:
\begin{equation}\label{3.19}
V(q_{min})=(\gamma_{g,q}N^{\frac{1}{3}} T^{2} \Lambda^4 / 2)^{1/4},
\end{equation}

where $N=(4/3 )[12 \pi / (33-2 n_{f})]$.  

So, the cut off in 
the model leads to finite integrals by avoiding 
the infra-red divergence, taking  
the consideration of the magnitude of $\Lambda$ and 
$T$ as same order of the lattice QCD as the characteristic feature of
the free energy assume to be in the same order/similar direction of other 
models.
 $g_{i}$ is degeneracy factor (color and particle-antiparticle degeneracy) 
which is $6$ for 
quarks and~$8$~for gluons and~$3$~for pions.
 The interfacial energy
obtained through a scalar 
Weyl-surface in Ramanathan et al.~[4,12] with  suitable 
modification to take care of 
the hydrodynamic effects is given as:  
\begin{equation}\label{3.22}
  F_{interface}= \frac{1}{4}\gamma R^{2}T^{3}. 
\end{equation}
This energy replaces the bag energy of MIT model as the MIT model 
produces drawback in the
numerical calculations of pressure and energy density.
The pion free energy is~[13]

\begin{equation}\label{3.25}
F_{\pi} = (3T/2\pi^2 )\nu \int_0^{\infty} q^2 dq \ln (1 - e^{-\sqrt{m_{\pi}^2 + q^2} / T}).
\end{equation}

Thus to calculate all these corresponding energies the
particle masses are taken as:
quark masses $m_u = m_d = 0 ~ MeV$ 
and $m_s = 0.15 ~ GeV$ just as 
taken in Ref.[11] and pion mass as $m_{\pi}=0.14~GeV$.

We can thus compute the total modified free energy $F_{total}$ as,
\begin{equation}
 F_{total}=\sum_{i} F_{i}~+~F_{interface}~+~F_{\pi},
\end{equation}
~ where $i$ stands for $u$,~$d$
and $s$~quark and gluon.   
\begin{figure}
\epsfig{figure=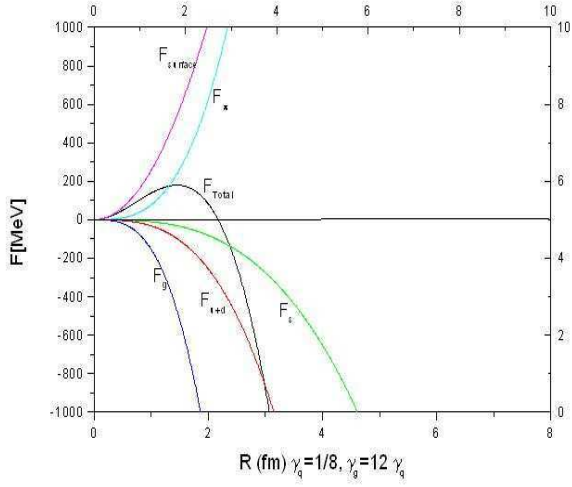,height=3.1in,width=3.2in}
\label{fig1}
\caption{\large  Individual free energy contribution~$F_{i}$~vs. R at~$\gamma_{q}=1/8~$, $\gamma_{g}=12\gamma_{q}$
at the particular temperature~$T=152~MeV$.}
\end{figure}
\begin{figure}
\epsfig{figure=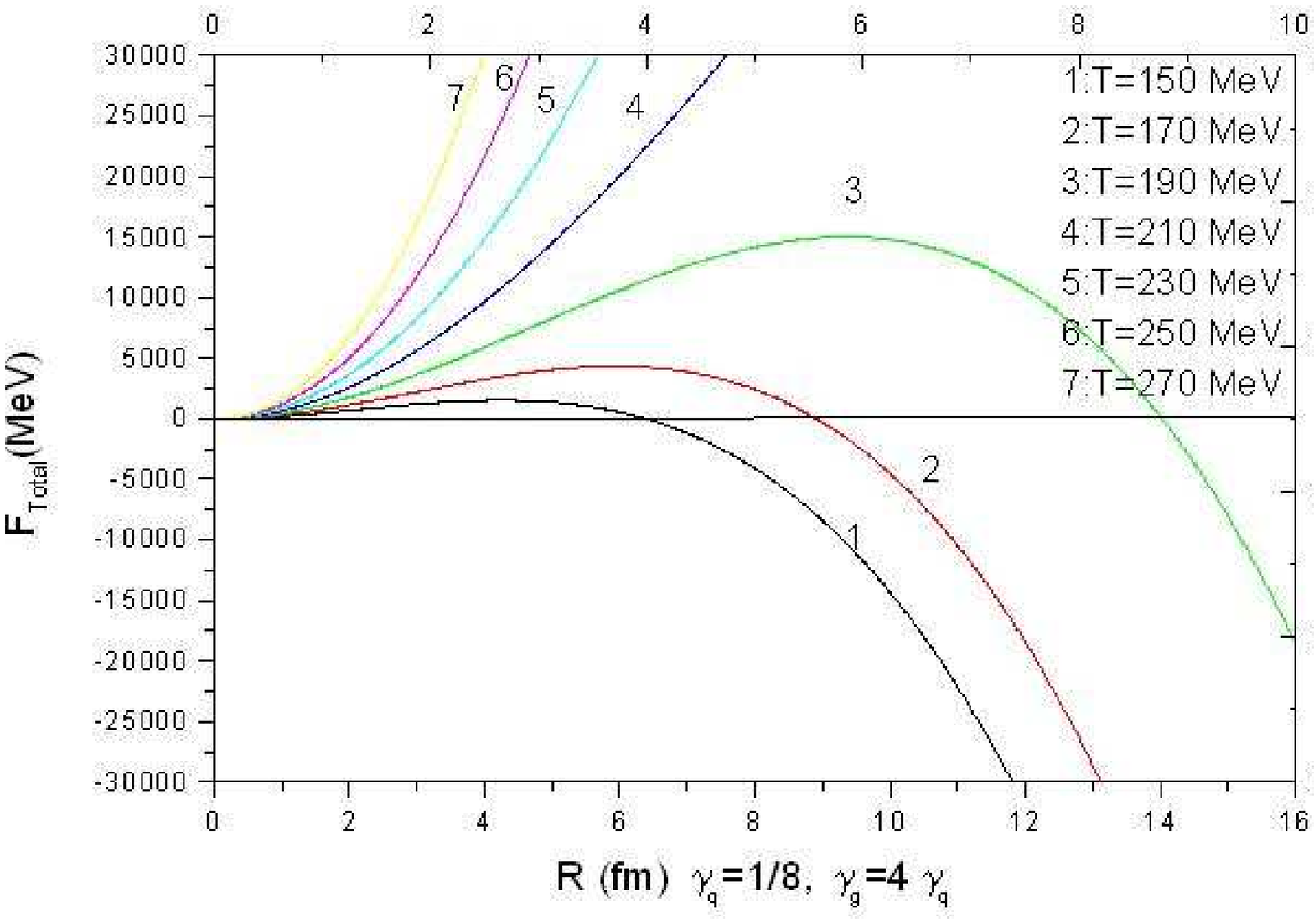,height=3.1in,width=3.2in}
\label{fig2}
\caption{\large  The free energy vs.~R~at~$\gamma_{q}=1/8~$, $\gamma_{g}=4\gamma_{q}$ for various values of temperature.}
\end{figure}
\begin{figure}
\epsfig{figure=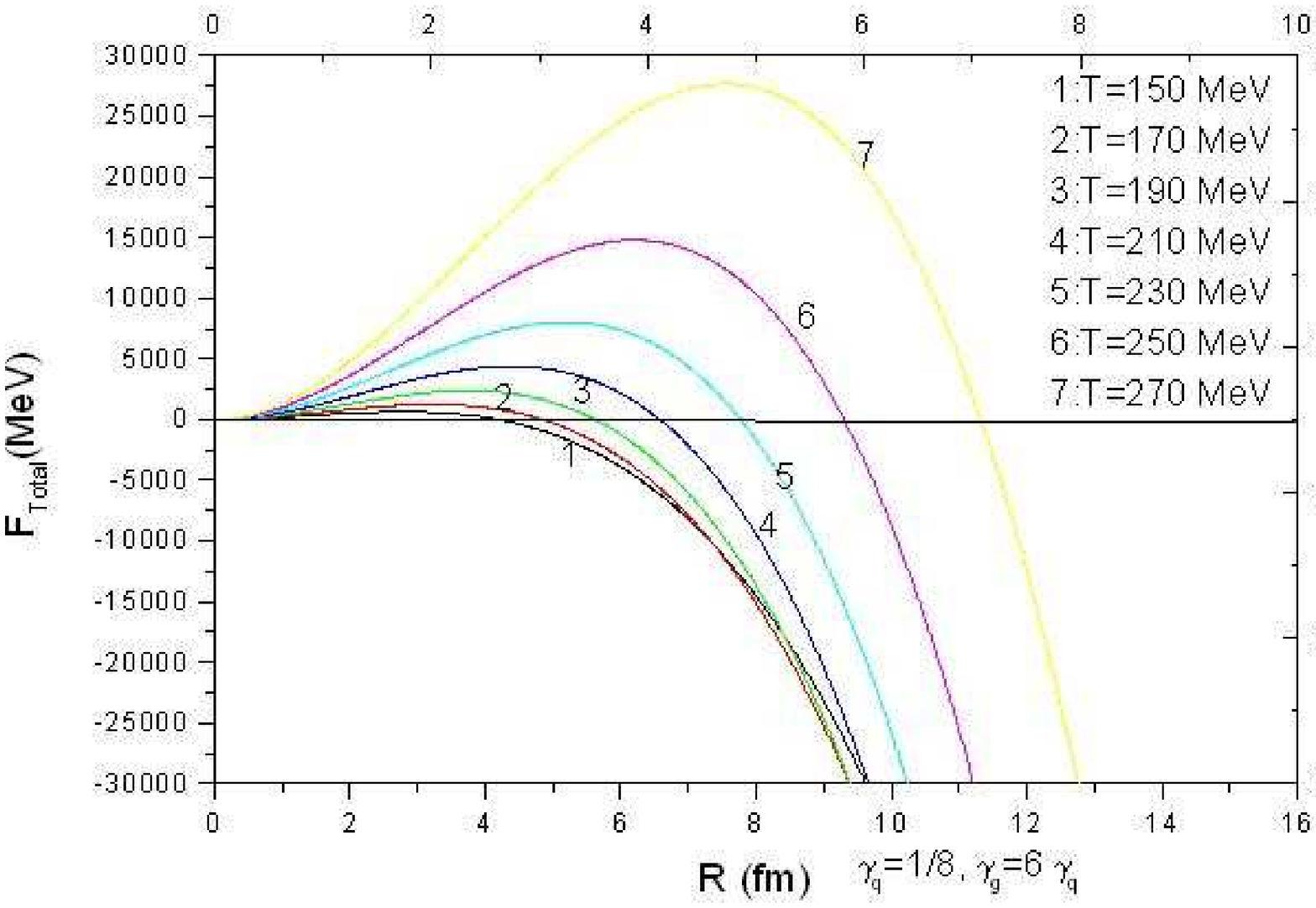,height=3.1in,width=3.2in}
\label{fig3}
\caption{\large  The free energy vs.~R~
at~$\gamma_{q}=1/8~$, $\gamma_{g}=6\gamma_{q}$ for the various values of temperature.}
\end{figure}
\begin{figure}
\epsfig{figure=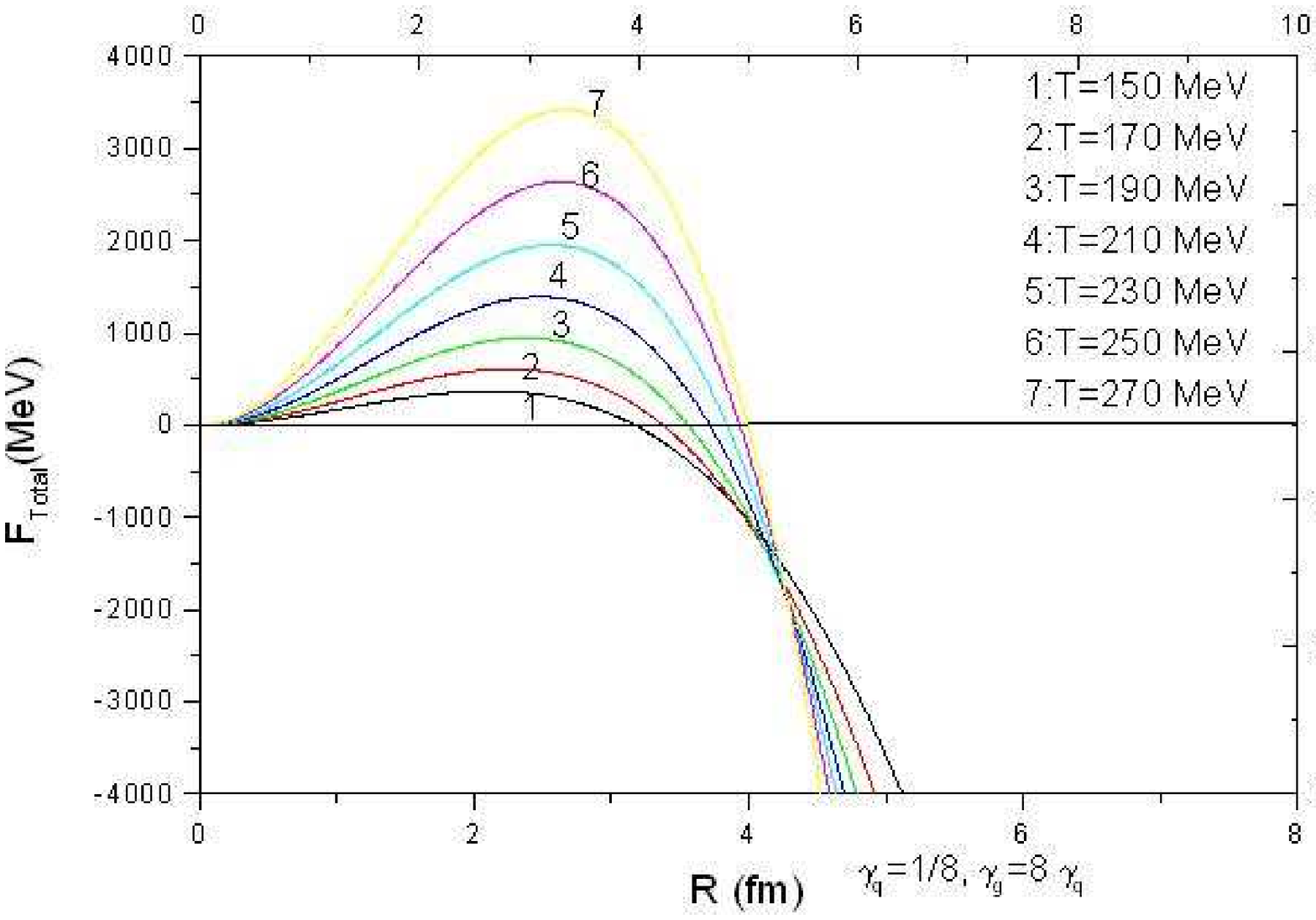,height=3.1in,width=3.2in}
\label{fig4}
\caption{\large  The free energy vs.~R~
at~$\gamma_{q}=1/8~$, $\gamma_{g}=8\gamma_{q}$ for the various values of temperature.}
\end{figure}
\begin{figure}
\epsfig{figure=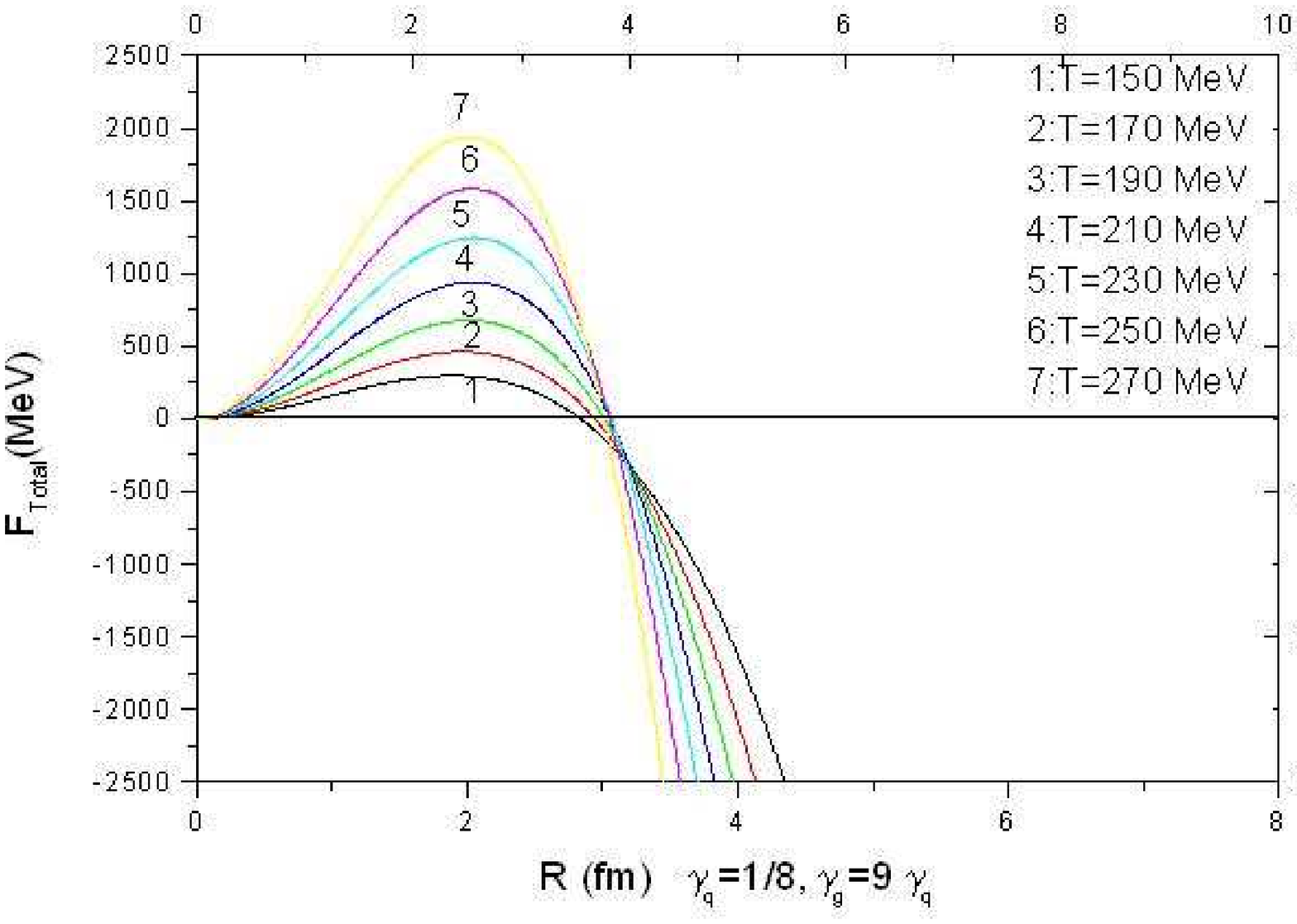,height=3.1in,width=3.2in}
\label{fig5}
\caption{\large  The free energy vs.~R~
at~$\gamma_{q}=1/8~$, $\gamma_{g}=9\gamma_{q}$ for the various values of temperature.}
\end{figure}
\begin{figure}
\epsfig{figure=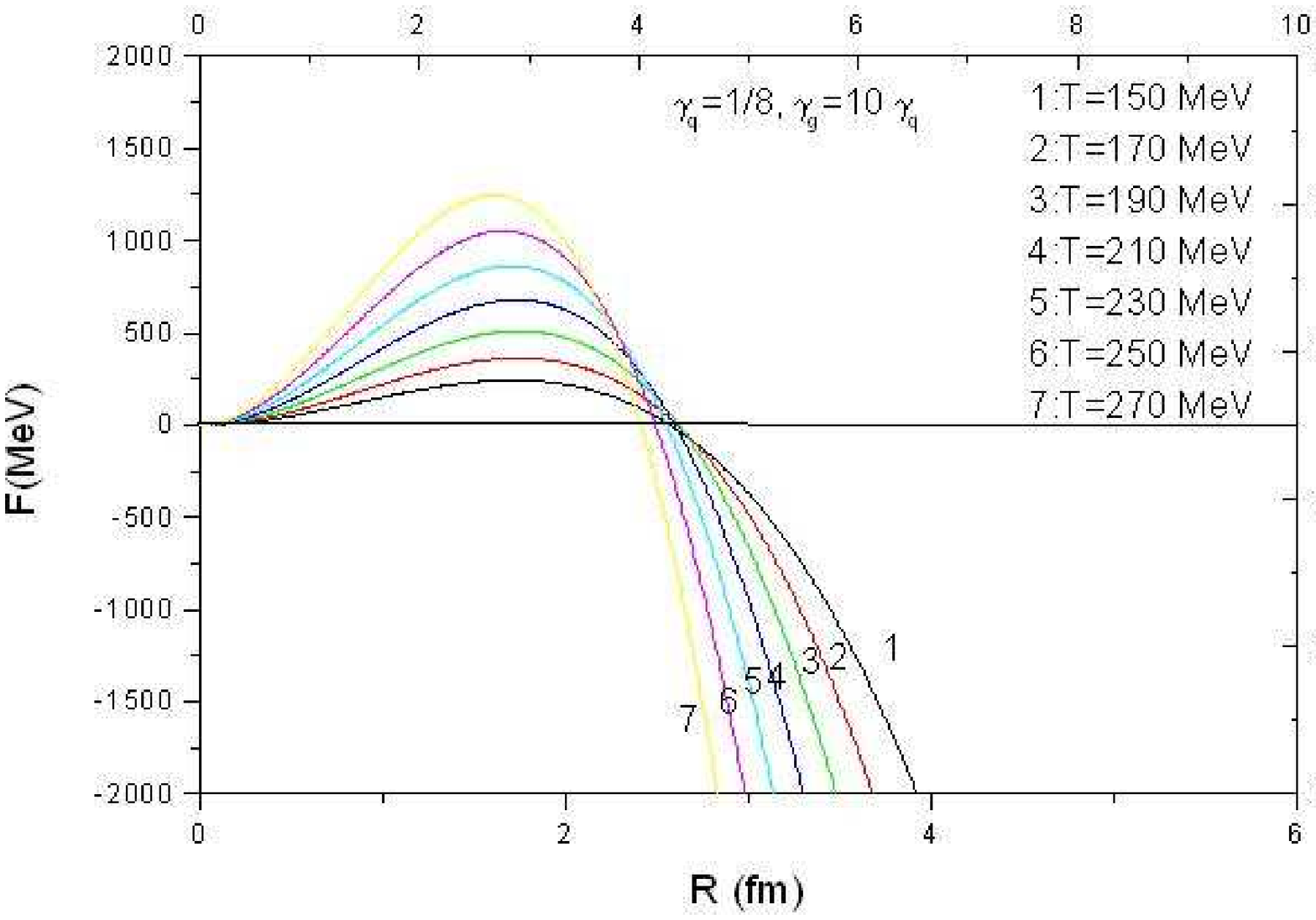,height=3.1in,width=3.2in}
\label{fig6}
\caption{\large  The free energy vs.~R~at~$\gamma_{q}=1/8~$, $\gamma_{g}=10\gamma_{q}~$for various values
 of temperature.}
\end{figure}
 %%%%%%%%%%%%%%%%%%%%%%%%%%%%%%%%%%%%%%%%%%%%%%%%%%%%%%%%%%%%%%%%%%%%%%%%%%%%%%%
\section{Results:} The free energy 
of the constituent particles of QGP fireball with the one loop 
correction factor in the peshier potential is numerically calculated. 
The evolution of QGP-hadron fireball
with the modification in the density of states of each particle 
is explained in the figures. The free energies of the individual
particles are shown in figure$~(1)~$ at a particular 
temperature $~T=152~MeV$ for the 
parametrization $~\gamma_{q}=1/8,~\gamma_{g}=12 \gamma_{q}$ and 
the energy shows
the behavior of QGP-hadron droplet formation.  
It means that with the inclusion
of one loop correction in the Peshier potential, the free energy of the system
is modified with the change in the amplitude
and stability of the droplet formation. So the flow parameters
have to change its value to reproduce the earlier results~[4] and 
the change in the 
parameter causes to increase the interaction between the constituent particles
showing the decrease in the amplitude of the free 
energy with these smaller flow parameters. 
\par The plots of various droplet formation 
for the various flow parameters ranging
from $4 \gamma_{q} \le \gamma_{g} \le 10 \gamma_{q}$ are shown in the 
figures~$(2-6)$.
In fig.$(2)$, there exists phase transition at the 
temperatures~$T=190~$MeV with flow parameters $ \gamma_{q} = 1/8,
\gamma_{g}= 4 \gamma_{q}$ and the changes of transition temperature
is also found upto the temperature $T=250~$MeV with the increase
of gluon parameter~$\gamma_{g} < 8 \gamma_{q}$.
In Fig.~$(3)$, it shows the unstable droplet formation with the
increase of $\gamma_{g}= 6 \gamma_{q}$ with the decrease of
amplitude with temperature. The formation of droplets in this range is 
also highly unstable till
the stable droplet formation.
In Figs.~$(4-6)$ we can easily observe the stability of droplet
formation with the flow parameters of $ \gamma_{q} = 1/8$ and
$~8 \gamma_{q}\le \gamma_{g}\le 10 \gamma_{q}$. 
The stability is obtained with the different
size of droplet and the stable droplets are found in the range
$~2.5-4.5~fm$ and its size decreases as the value of the gluon 
parameter increases. This is obtained with the decrease in the
quark and gluon flow parameters.
\section{Conclusion:} 
We can conclude from these results that due to the presence of loop
correction in the potential the system increase the stability of droplet 
formation and decrease in the amplitude of free energy with the new flow
parameters. So, we can further
study the surface tension and thermodynamic properties of QGP on the
basis of these smaller droplet size. This means that the results with 
one loop correction in the peshier potential make the ad-hoc choice
of flow parameters most appropriate and will be in agreement 
with lattice gauge
expectation values. This is good outputs in the free energy  with smaller
quark and gluon parameters over free
energy of earlier results.  
\\   
\\
{\bf Acknowledgments:}
\\
\\
We thank K. K. Gupta for useful discussions
and critical reading of the manuscript and author (SSS) thanks 
the University 
for providing research and development
strengthen grants.  
\\
\\
  
{\bf References :}

\begin{enumerate}
\item{A. Ali Khan {\it et al}., CP-PACS Collab: Phys. Rev D {\bf 63}, 
034502 (2001);
E. Karsch, A. Peikert and E. Laermann, Nucl. Phys. B {\bf 605}, 579 (2001)}.
\item{H. Satz,{ \it CERN-TH-2590}, 18pp (1978); F. Karsch, E. Laermann, A. Peikert, Ch. Schmidt and S. Stickan, Nucl. Phys. B {\bf 94}, 411 (2001); F. Karsch and H. Satz, Nucl. Phys. A {\bf 702}, 373 (2002).}
\item{A. Peshier, B. K$\ddot{a}$mpfer, O. P. Pavlenko and G. Soff, 
Phys. Lett. B {\bf 337}, 235 (1994); V. Goloviznin and H. Satz, Z. Phys. C {\bf 57}, 671 (1993).}
\item{R. Ramanathan, Y. K. Mathur, K. K. Gupta and A. K. Jha, Phys. Rev. C {\bf 70}, 027903 (2004); R. Ramanathan, K. K. Gupta, A. K. Jha and S. S. Singh, Pram. J. Phys. {\bf 68}, 757 (2007).}
\item{S. S. Singh, D. S. Gosain, Y. Kumar and A. K. Jha, Pram. J. Phys. {\bf 74}, 27 (2010).}
\item{ N. Brambilla, A. Pineda, J. Soto and A. Vairo, Phys. Rev D {\bf 63}, 014023 (2001).}
\item{K. Melnikov and A. Yelkhovsky, Nucl. Phys. B {\bf 528}, 59 (1998);
A. H. Hoang, Phys. Rev D {\bf 59}, 014039 (1999).}
\item{W. Fischler, Nucl. Phys. B {\bf 129}, 157 (1977); A. Billoire, Phys. Lett B {\bf 92}, 343 (1980).}
\item{A. V. Smirnov, V. A. Smirnov, M. Steinhauser, 
Phys. Lett. B {\bf 668}, 293 (2008);
A. V. Smirnov, V. A. Smirnov and M. Steinhauser, 
Phys. Rev. Lett. {\bf 104}, 112002 (2010).}
\item{A. D. Linde, Nucl. Phys. B {\bf 216}, 421 (1983); E. Fermi, Zeit F. Physik {\bf 48}, 73 (1928); L. H. Thomas , Proc. Camb. Phil. Soc. {\bf 23}, 542 (1927); H. A. Bethe, Rev. Mod. Phys. {\bf 9}, 69 (1937).}
\item{G. Neergaad and J. Madsen, Phys. Rev. D {\bf 60}, 05404 (1999);
M. B. Christiansen and J. Madsen, J. Phys. G {\bf 23}, 2039 (1997).}
\item{H. Weyl, \it {Nachr. Akad. Wiss Gottingen} 110 (1911). }
\item{R. Balian and C. Block, Ann. Phys. (NY) {\bf 60}, 401 (1970).}
\end{enumerate}
\end{document}